\documentclass[twocolumn,aps,prl,superscriptaddress,10pt]{revtex4-1}
\usepackage{graphicx,amsmath}
\newcommand{\expval}[1]{\langle #1\rangle}

\newcommand{\RM}[1]{\MakeUppercase{\romannumeral #1{}}}
\begin{document}
%-------------------------------------------------------------------------
\title{Dipole coupling of a double quantum dot to a microwave resonator}
%-------------------------------------------------------------------------
\author{T.~Frey}
\affiliation{Department of Physics, ETH Zurich, CH-8093, Zurich, Switzerland.}
\author{P.~J.~Leek}
\affiliation{Department of Physics, ETH Zurich, CH-8093, Zurich, Switzerland.}
\author{M.~Beck}
\affiliation{Department of Physics, ETH Zurich, CH-8093, Zurich, Switzerland.}
\author{A.~Blais}
\affiliation{D\'{e}partement de Physique, Universit\'{e} de Sherbrooke, Sherbrooke, Qu\'{e}bec, J1K 2R1 Canada.}
\author{T.~Ihn}
\affiliation{Department of Physics, ETH Zurich, CH-8093, Zurich, Switzerland.}
\author{K.~Ensslin}
\affiliation{Department of Physics, ETH Zurich, CH-8093, Zurich, Switzerland.}
\author{A.~Wallraff}
\email{andreas.wallraff@phys.ethz.ch}
\affiliation{Department of Physics, ETH Zurich, CH-8093, Zurich, Switzerland.}
%%-------------------------------------------------------------------------
\date{\today}
\maketitle
%-------------------------------------------------------------------------

%-------------------------------------------------------------------------
% Introduction
%-------------------------------------------------------------------------
\textbf{Quantum coherence in solid-state systems has been demonstrated in superconducting circuits \cite{Nakamura1999} and in semiconductor quantum dots \cite{Hayashi2003}. This has paved the way to investigate solid-state systems for quantum information processing with the potential benefit of scalability compared to other systems based on atoms, ions and photons \cite{natureinsight2008}. Coherent coupling of superconducting circuits to microwave photons, circuit quantum electrodynamics (QED) \cite{Wallraff2004}, has opened up new research directions \cite{Schoelkopf2008} and enabled long distance coupling of qubits \cite{Majer2007}. Here we demonstrate how the electromagnetic field of a superconducting microwave resonator can be coupled to a semiconductor double quantum dot. The charge stability diagram of the double dot, typically measured by direct current (DC) transport techniques \cite{kouwenhoven1997}, is investigated via dispersive frequency shifts of the coupled resonator. This hybrid all-solid-state approach offers the potential to coherently couple multiple quantum dot and superconducting qubits together on one chip, and offers a method for high resolution spectroscopy of semiconductor quantum structures.}

Semiconductor quantum dots are highly controllable solid-state quantum systems \cite{vanderWiel2003,Hanson2007}. Charge measurements in the radio frequency (RF) regime have been demonstrated \cite{Lu2003,Reilly2007,Cassidy2007,Mueller2010} and recently a lumped element RF resonator was used to measure the quantum capacitance of a double dot \cite{Petersson2010}. A number of schemes have been proposed for the scaling of quantum dot based quantum information processing \cite{Childress2004,Taylor2005,Burkard2006,Taylor2006}. In this work we implement a form of circuit QED \cite{Wallraff2004}, coupling charge states of a double dot to the field of an on-chip microwave transmission line resonator \cite{Childress2004}.

%-------------------------------------------------
%part 1 describe sample + measurement setup
%-------------------------------------------------
\begin{figure*}
\includegraphics[width=1.4 \columnwidth]{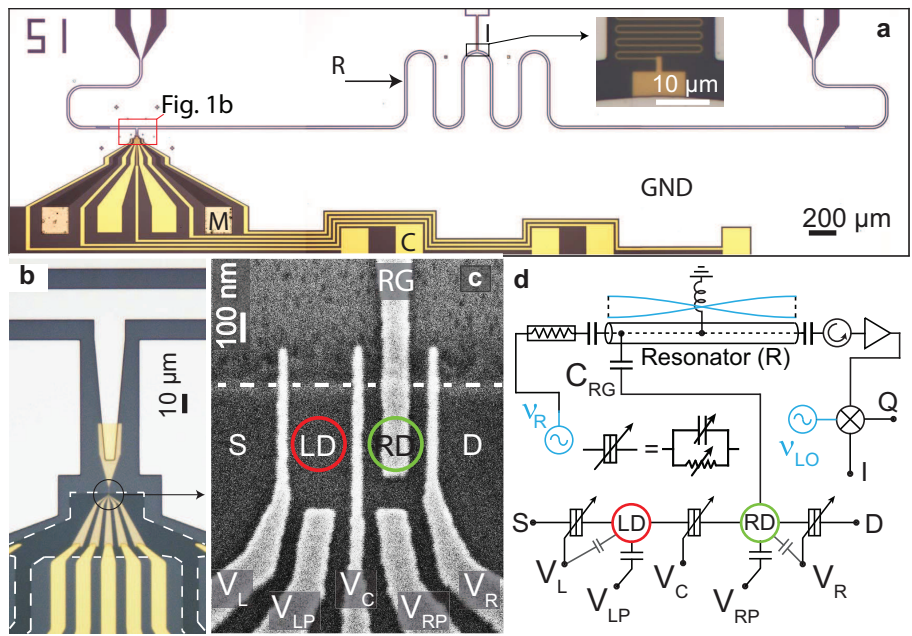}
\caption{\textbf{Images and schematic of the hybrid quantum device.} \textbf{a}, Optical micrograph of the microwave resonator (R), with integrated double quantum dot, ohmic contacts (M), top gates (C), ground plane (GND, on chip inductor (I). Inset: magnified view of inductor (I). \textbf{b}, Enlarged view of the device near the double quantum dot. The mesa edge is highlighted with a dashed line. \textbf{c}, Scanning electron micrograph of the gate structure defining the double quantum dot (LD, RD). RG marks the gate connected to the resonator, ($V_{\rm{L}}$, $V_{\rm{LP}}$, $V_{\rm{C}}$, $V_{\rm{RP}}$, $V_{\rm{R}}$) label top gate voltages, and S, D, the 2DEG source and drain. \textbf{d}, Electric circuit representation of the double quantum dot coupled to the resonator. The double quantum dot is tuned with voltages $V_{\rm{L}}$, $V_{\rm{LP}}$, $V_{\rm{C}}$, $V_{\rm{RP}}$, $V_{\rm{R}}$, and connected to the resonator via the capacitance $C_{\rm{RG}}$. The resonator is driven with a microwave signal at frequency $\nu_{\rm{R}}$. The transmitted signal passes through a circulator, is amplified and mixed with the local oscillator $\nu_{\rm{LO}}$ to obtain the field quadratures $I$ and $Q$.}
\label{one}
\end{figure*}

The sample investigated is shown in Fig.~\ref{one}a-c along with an electrical circuit schematic (Fig.~\ref{one}d).
The microwave resonator (see Fig.~\ref{one}a) is realized using a $200~\rm{nm}$ thick Aluminum coplanar waveguide on GaAs and is capacitively coupled to an input and output line to probe its transmission spectrum. The double quantum dot (see Fig.~\ref{one}c), is fabricated at the position of an anti-node of the standing wave field distribution of the resonator. The left and right dots (LD, RD) are arranged in series with respect to the source and drain (S, D) and are realized on an $\rm{Al_{x}Ga_{1-x}As}$ heterostructure with the two-dimensional electron gas (2DEG) at a depth of about $35~\rm{nm}$ below the surface.

To enable a strong coupling between the two systems, an additional gate (RG) (Fig.~\ref{one}c) was implemented, which extends from the resonator to the right quantum dot. This gives a selective capacitive coupling of the resonator to the right quantum dot, confirmed by DC biasing the resonator via an on-chip inductor (Fig.~\ref{one}a, inset). This results in a strong dipole coupling of the resonator to two charge states in which an electron is on either the left or right quantum dot. In order to accommodate the gate (RG), a design is realized in which the dots are placed at the mesa edge (beyond which the 2DEG is etched away), which is used as part of the confining potential. To complete the formation of the dots, negative voltages are applied to metallic top gates (Fig.~\ref{one}a), below which the 2DEG is then depleted.

The static potential on the dots is tuned via the two plunger gates $V_{\rm{LP}}$ and $V_{\rm{RP}}$. To allow electron transport, the two dots are connected to each other and to the source (S) and drain (D) contacts through tunnel barriers, tunable by $V_{\rm{L}}$, $V_{\rm{C}}$ and  $V_{\rm{R}}$ (see Fig~\ref{one}c, d). Due to finite capacitive coupling, these tunnel barrier gates also tune the dot potentials in a similar way to the plunger gates. The resonator (R) is probed with a coherent microwave signal at a frequency $\nu_{\rm{R}}$. The amplitude $A$ and phase $\phi$ of the transmitted signal are extracted from the field quadratures $I$ and $Q$, as $Ae^{i\phi}=I+iQ$, measured by heterodyne detection \cite{Wallraff2004}. The experiments are performed in a dilution refrigerator with a base temperature of $T \approx 10~\rm{mK}$.

\begin{figure}
\centering
\includegraphics[width=1 \columnwidth]{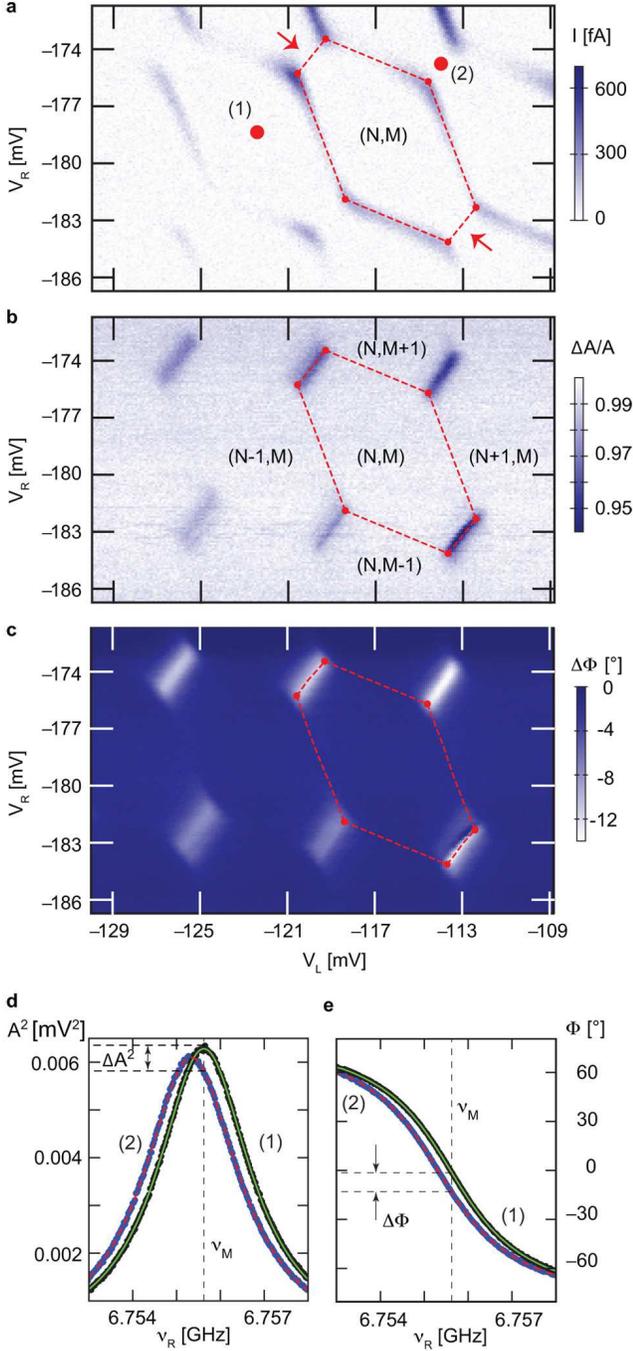}
\caption{\textbf{Double quantum dot measurements.} \textbf{a}, Measurement of the direct current through the double quantum dot versus $V_{\rm{L}}$, $V_{\rm{R}}$ (no microwave signal applied). A small background current is subtracted from each horizontal line of the dataset. The dashed line outlines a region with fixed electron number (N, M). \textbf{b}, Resonator transmission amplitude at fixed measurement frequency $\nu_{\rm{M}}$ in the same gate voltage range as in \textbf{a}. The red dashed lines highlight the same hexagon as in \textbf{a}. \textbf{c}, Transmission phase change with respect to $\nu_{\rm{M}}$, gate voltages as in \textbf{a}. Red lines as in \textbf{b}.
\textbf{d}, Measured resonator transmission spectra of the microwave resonator measured at position (1) (black points) and (2) (blue points), indicated in \textbf{a} with fitted solid (green) and dashed (red) line. \textbf{e}, Measured phase (points) with fits as in \textbf{d}.
}\label{two}
\end{figure}

We first investigated the DC transport properties of the double quantum dot (see Fig.~\ref{two}a) with a source-drain voltage of $V_{\rm{SD}} = 50~\rm{\mu V}$ applied. To change the electrostatic potentials on the quantum dots, the voltages $V_{\rm{L}}$ and $V_{\rm{R}}$ are varied while all the other gate voltages are fixed. A hexagon charge stability pattern is observed, typical for electron transport through double quantum dots \cite{vanderWiel2003}. Within each hexagon, the number of electrons  (N, M) in the dots is fixed. At the triple points, where three charge states are degenerate, charge transport is possible and a conductance resonance is observed. The current seen along four of the hexagon boundaries is caused by co-tunneling or molecular orbital states \cite{Gustavsson2008}. Along the two remaining hexagon boundaries (indicated with arrows in Fig.~\ref{two}a) where no DC current was measured, two quantum dot charge states with the same total number of electrons are degenerate \cite{vanderWiel2003}. The charging energies $E_{\rm{C}}$ of both quantum dots, extracted from Coulomb diamond measurements, are $E_{\rm{C}} \approx 1~\textrm{meV}\approx h\times240~\textrm{GHz}$. Both quantum dots contain on the order of 100 electrons and an electron temperature of $T_{e}\lesssim 135~\rm{mK}$ is estimated from Coulomb resonance linewidths \cite{kouwenhoven1997}.

%-------------------------------------------------
%part 2 show double quantum dot features + equivalent RF measurements
%-------------------------------------------------

The microwave resonator is designed as in Ref.~\cite{Frey2011} and we measure a fundamental frequency $\nu_0\approx 6.7549~\rm{GHz}$ and loaded quality factor \cite{Goeppl2008} $Q_{L}\approx 2630$, with all gates grounded such that no quantum dots are formed. The resonator is well isolated from thermal radiation originating from the higher temperature stages of the dilution refrigerator, and is hence approximately in its thermal ground state with average thermal photon number $n<1$.

When the double quantum dot is formed by applying appropriate gate voltages, we find that the resonator frequency and linewidth are sensitively dependent on the double dot gate voltage configuration. In Fig.~\ref{two}d and e, the amplitude and phase transmission spectrum of the microwave resonator is shown, for the two voltage settings indicated with (1) and (2) in Fig.~\ref{two}a. Note that here, and in all further microwave transmission measurements, source and drain are grounded. In both configurations (1) and (2), transport is blocked between the leads due to Coulomb blockade. However, in (1), the electron number in both dots is fixed, whereas in (2), a pair of left and right dot charge states are degenerate and hybridized states are formed. Different resonance frequencies and maximum transmission amplitudes are seen in the two cases. This indicates that there is a strong interaction between the double dot and the resonator.

We now proceed to measure the double dot using the resonator, working with a fixed probe frequency $\nu_{\rm{M}}$ and recording the transmitted amplitude and phase. Such measurements for the same gate voltage ranges as in Fig.~\ref{two}a are shown in Fig.~\ref{two}b and c, and the same hexagon pattern shown in Fig.~\ref{two}a is also overlayed (red dashed line). Amplitude and phase changes are observed at the triple points and along the interdot charge transfer lines where left and right dot charge states are degenerate. There is however no clear change of the microwave signal measured along the cotunneling lines. The hexagon pattern shown in Fig.~\ref{two}b and c could be observed over a range of more than 10 electrons in both left and right dot, including gate voltage settings where the tunneling rates were so small that the DC current through the double quantum dot could not be detected in direct transport measurements ($I\lesssim 80~\rm{fA}$). These measurements suggest that the microwave signal is sensitive to electron exchange between the two dots rather than with the leads.

\begin{figure}[bp]
\begin{center}
\includegraphics[width=1 \columnwidth]{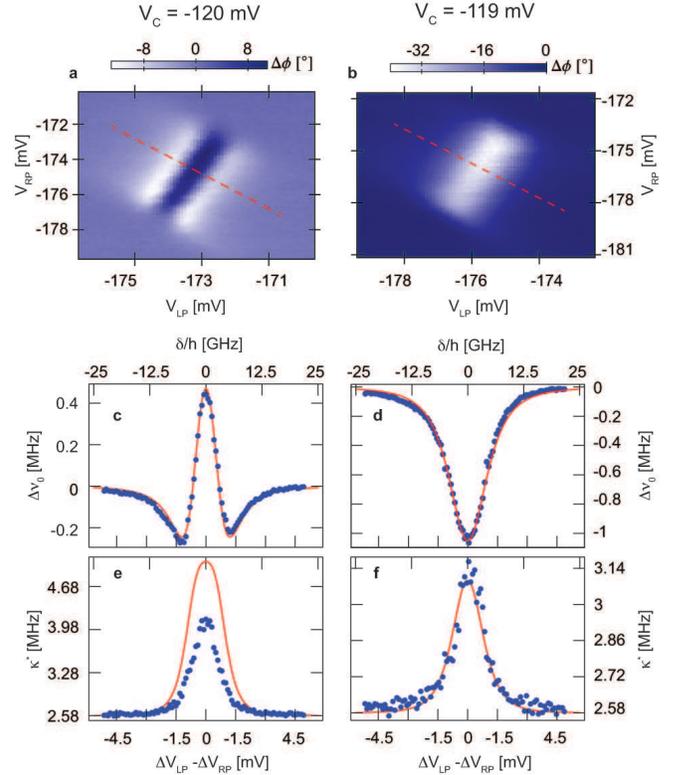}
\caption{\textbf{Detailed resonator measurements at an interdot charge transfer line.}
\textbf{a}, Transmission phase at $\nu_{\rm{M}}$ as a function of $\rm{V_{LP}}$ and $\rm{V_{RP}}$ for $\rm{V_{C}}=-120~\rm{mV}$. Red dashed line indicates axis swept in \text{c,e}.
\textbf{b}, Transmission phase measured for $\rm{V_{C}}=-119~\rm{mV}$. Red dashed line indicates axis swept in \text{d,f}.
\textbf{c,d}, Measured resonator frequency shift (blue points) along the axes shown in \textbf{a} and \textbf{b} respectively.
\textbf{e,f}, Measured resonator linewidth $\kappa^{*}$ (blue points) along the axes shown in \textbf{a} and \textbf{b} respectively.
Solid red lines in \textbf{c-f} are numerical simulations.
}\label{three}
\end{center}
\end{figure}

We now examine the transmitted phase at $\nu_{\rm{M}}$ in the vicinity of one particular interdot charge transfer line, for different values of interdot tunnel coupling energy $t$, tuned using $V_{\rm{C}}$. More negative $V_{\rm{C}}$ gives a smaller $t$. The two plunger gates $V_{\rm{LP,RP}}$ are swept (rather than $V_{\rm{L,R}}$), to minimize the change of the tunnel rates to the leads.
In Fig.~\ref{three}a and b the center gate is set to $V_{\rm{C}}=-120~\rm{mV}$ and $V_{\rm{C}}=-119~\rm{mV}$ respectively. $V_{\rm{L,R}}$ are set such that the tunneling rates to the leads are too small to measure DC transport. Despite the small change in $V_{\rm{C}}$ between the two measurements, a clear difference is observed. In Fig.~\ref{three}a, two regions of negative phase shift are seen either side of a region of positive phase shift along the interdot charge transfer line. In Fig.~\ref{three}b however, a single region of negative phase shift is seen, similar to the features shown in Fig.~\ref{two}c. Note that negative/positive phase shifts translate directly to negative/positive resonance frequency shifts (see Fig.~\ref{two}d and e).

%-------------------------------------------------
%part 3 the data of the crossings
%-------------------------------------------------

In order to directly determine the resonator frequency shift $\Delta \nu_{\rm{0}}$ and linewidth $\kappa^{*}$ across the interdot charge transfer line, $V_{\rm{LP}}$ and $V_{\rm{RP}}$ are swept in parallel following the red dashed lines in Fig.~\ref{three}a and b. For each voltage setting the full transmission spectrum (see Fig.~\ref{two}d) is recorded, which allows pure frequency shifts to be easily distinguished from dissipative effects. The parameters $\Delta \nu_{0}$ and $\kappa^{*}$ are obtained by fitting these spectra to Lorentzians, and are plotted with respect to the plunger gate voltage changes for the two different interdot tunnel coupling strengths in Fig.~\ref{three} c-f.

In order to explain the observations seen in Fig.~\ref{three}, the hybrid system is modeled as a charge qubit \cite{Hayashi2003} coupled to a quantized harmonic oscillator, with the Jaynes-Cummings Hamiltonian in the rotating wave approximation \cite{Childress2004},
%-------------------------------------------------
%part 4 explain cavity QED model briefly
%-------------------------------------------------
\begin{eqnarray}
H = h\nu_0(\hat{n}+\frac{1}{2}) + \frac{h\nu_{\rm{q}}}{2}\hat\sigma_{\rm{z}}+ \hbar g\sin\theta(\hat{a}^{\dagger}\hat\sigma^{-} +\hat{a}\hat\sigma^{+}),\label{modeleq} \\
\nu_{\rm q}=\sqrt{\delta^2+4t^2},~~~~\sin\theta=2t/\sqrt{\delta^2+4t^2}.~~~~~~~~\nonumber
\end{eqnarray}
Here $\nu_0$ is the resonator frequency, $\delta$ the detuning energy between the charge qubit states, $t$ the tunnel coupling energy between the dots, and $\hbar g$ the coupling energy between the resonator and the qubit. The photon number operator is $\hat{n}=a^{\dagger}a$, and $\sigma^{+/-}$ and $\sigma_{z}$ are Pauli operators for the qubit.

\begin{figure}[tp]
\centering
\includegraphics[width=1 \columnwidth]{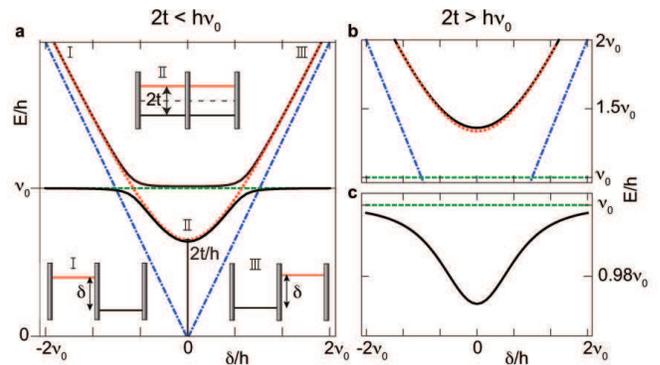}
\caption{\textbf{Eigenenergies of a charge qubit coupled to a resonator.} \textbf{a}, Schematic of the two lowest eigenenergies of the coupled system for $2t<h\nu_{0}$. Horizontal green dashed line indicates the bare resonator frequency $\nu_{0}$. Red dotted line shows the bare transition frequency $\nu_{\rm{q}}$ between the double dot charge states. The two sloped blue dashed-dotted lines indicate the detuning $\delta$. Insets \RM{1}, \RM{3}: Schematic of the two double dot charge states being detuned by energy $\delta$, in the limit $\delta\gg t$. Inset \RM{2}: Schematic of the hybridized charge states in the double quantum dot, split by $2t$ at $\delta=0$. \textbf{b}, Transition energy of the charge qubit for $2t>h\nu_{0}$ in the presence of the nearby resonator. Dashed and dotted lines as in \textbf{a}. \textbf{c}, Resonator frequency in the presence of the nearby charge qubit transition for the same case as \textbf{b}.}\label{four}
\end{figure}

The two lowest eigenenergies of the coupled system are displayed as a function of $\delta$ in Fig.~\ref{four}a, and Fig.~\ref{four}b and c, for the cases $2t<h\nu_{0}$ and $2t>h\nu_{0}$ respectively. The bare resonator frequency $\nu_0$ and bare charge qubit transition frequency $\nu_{\rm{q}}$ are modified in the presence of a finite coupling $g$ to give the solid black lines in Fig.~\ref{four}a. Within this model, which does not include decoherence, an avoided level crossing occurs at $\nu_{\rm{q}}=\nu_0$ \cite{Blais2004}. In Fig.~\ref{four}b and c, the dispersive case is displayed in which $\nu_{\rm{q}}$ is far detuned from $\nu_0$, for all $\delta$. The presence of the qubit (Fig.~\ref{four}b) now only results in a shift of the resonator frequency to values below $\nu_0$ (Fig.~\ref{four}c) \cite{Blais2004}.

We now compare the data shown in Fig.~\ref{three}c-f to steady-state numerical simulations of the Hamiltonian (equation~(\ref{modeleq})), including cavity and qubit damping, and qubit decoherence, in a Markovian master equation formulation \cite{WallsMilburnBook}. The results are shown as red lines in Fig.~\ref{three}c-f. We take a relaxation rate $\gamma_1/2\pi=100~\rm{MHz}$ typical for charge qubits \cite{Fujisawa1998}, and the bare resonator frequency $\nu_0=6.75564~\rm{GHz}$ and linewidth $\kappa/2\pi=2.58~\rm{MHz}$ from measurements. A factor is used to convert gate voltage settings to detuning energy $\delta$, shown on the top axis of Fig.~\ref{three}c-f. We find that the simulations lie close to the data if we include a coupling strength $g/2\pi=50~\rm{MHz}$ for all datasets, but take a different tunneling energy $t$ and dephasing rate $\gamma_{\rm{\phi}}$ for the two different settings of $V_{\rm{C}}$. For $V_{\rm{C}}=-119~\rm{mV}$ (Fig.~\ref{three}d and f), we use $2t/h=9~\rm{GHz}$, and $\gamma_\phi/2\pi=0.9~\rm{GHz}$, corresponding to the dispersive case (Fig.~\ref{four}c), to obtain close agreement with the phase shift and linewidth data. For $V_{\rm{C}}=-120~\rm{mV}$ (Fig.~\ref{three}c and e), we use $2t/h=6.1~\rm{GHz}$, and a significantly larger $\gamma_\phi/2\pi=3.3~\rm{GHz}$. In this case, the charge qubit crosses the resonator in the model, but anticrossings are not observed due to the large dephasing $\gamma_\phi$. Moreover, while the phase shift is in close agreement, the measured linewidth (Fig.~\ref{four}e) is lower than in the simulation. There are many possible causes for this discrepancy. Indeed, this simple model does not take into account the $\delta$-dependence of the double quantum dot relaxation \cite{Fujisawa1998} and dephasing \cite{Hayashi2003,Petersson2010a} rates. Moreover, our Markovian approach assumes white rather than 1/f noise. This can be expected to lead to discrepancies, especially given the large value of the dephasing rate used here.

The result of this comparison to simulation indicates that the qubit dephasing is by far the dominant decoherence mechanism, and it was checked that inclusion of finite temperature or a higher value of $\gamma_1$ could not explain the data. It was also confirmed both experimentally and in simulation that the results in Fig.~\ref{three} are independent of coherent resonator population up to $\expval{n_{coh}}\sim10$. The displayed measurements used $\expval{n_{coh}}\sim10$ to obtain good signal-to-noise while keeping the measurement time short to avoid rare charge reconfigurations.

We have successfully demonstrated the dipole coupling of a double quantum dot to an on-chip superconducting microwave resonator, and probed the double dot charge stability diagram by measuring resonator frequency shifts. Two different characteristic regimes with interdot tunnel couplings of the double quantum dot above or below the resonator frequency could be observed and explained. Our architecture offers a new way to probe semiconductor quantum systems in the microwave regime, and may be used, for example, for high energy-resolution measurements of double quantum dots \cite{Jin2011}, and fast time-resolved measurements, in addition to being a promising platform for scalable hybrid solid-state quantum information processing. The presented scheme could be extended to other material systems, manipulating and reading out spin qubits \cite{Hanson2007} and coupling them to a microwave resonator using either ferromagnetic leads \cite{Cottet2010} or spin orbit effects \cite{Trif2008}.

While preparing this manuscript we became aware of a related work \cite{Delbecq2011}, in which a carbon nanotube single quantum dot and its fermionic leads were coupled to a microwave resonator.
%-------------------------------------------------------------------------
\vspace*{-5mm}
\bibliographystyle{apsrev}

~

\textbf{Acknowledgements}\\
We thank C.~R\"ossler, T.~M\"uller and D.~Loss for valuable discussions, C.~Lang for measurement software, P.~Studerus for excellent technical support, and T.~Schoch and V.~Tshitoyan for their contributions to mesa-edge dot development and numerical simulations respectively.
This work was supported financially by EU IP SOLID, by the Swiss National Science Foundation through the National Center of Competence in Research `Quantum Science and Technology' and by ETH Zurich. A.~B.~was supported by NSERC, the Alfred P.~Sloan Foundation, and CIFAR.\\

\textbf{Author contributions}\\
T.F. and P.J.L. fabricated the sample, and set up and carried out the experiments. P.J.L. and A.B. carried out the simulations. M.B. carried out the molecular-beam-epitaxial growth of the GaAs heterostructure. T.F., P.J.L. and K.E. wrote the manuscript in discussion with all authors. T.I., K.E. and A.W. supervised and managed the project.

\end{document}